# IRAS-based Whole-Sky Upper Limit on Dyson Spheres

Richard A. Carrigan, Jr.,
Fermi National Accelerator Laboratory, Batavia, IL 60510, USA
carrigan@fnal.gov
630-840-8755

ABSTRACT

A Dyson Sphere is a hypothetical construct of a star purposely cloaked by a thick swarm of broken-up planetary material to better utilize all of the stellar energy. A clean Dyson Sphere identification would give a significant signature for intelligence at work. A search for Dyson Spheres has been carried out using the 250,000 source database of the IRAS infrared satellite which covered 96% of the sky. The search has used the Calgary database for the IRAS Low Resolution Spectrometer (LRS) to look for fits to blackbody spectra. Searches have been conducted for both pure (fully cloaked) and partial Dyson Spheres in the blackbody temperature region $100 \leq T \leq 600$ °K. When other stellar signatures that resemble a Dyson Sphere are used to eliminate sources that mimic Dyson Spheres very few candidates remain and even these are ambiguous. Upper limits are presented for both pure and partial Dyson Spheres. The sensitivity of the LRS was enough to find Dyson Spheres with the luminosity of the sun out to 300 pc, a reach that encompasses a million solar- type stars.

*Subject headings:* circumstellar matter - extraterrestrial intelligence - infrared radiation – stars: AGB and post-AGB – stars: carbon

ACKNOWLEDGEMENT

Operated by the Fermi Research Alliance, LLC under Contract No. DE-AC02-07CH11359 with the United States Department of Energy.

1. INTRODUCTION

Advanced civilizations might undertake astroengineering projects on heroic scales. Searches for signs of astroengineering to look for intelligence elsewhere in the Universe represent an approach that is complementary to conventional radio and optical SETI. A Dyson Sphere (Dyson 1960) is a good example of an astroengineering project or "cosmic archaeology" artifact.

In 1960 Dyson suggested that an advanced civilization might break up a star's planets into very small planetoids or pebble-sized fragments to form a "loose collection or swarm of objects" that would gather all the visible light coming from the star. The shroud of objects forming a spherical shell would greatly increase the useful area and available energy for advanced activities. A Dyson Sphere like this that totally envelopes the host star is called a pure Dyson Sphere to distinguish it from a partial Dyson Sphere that does not fully cloak the star.

Unlike SETI signals (Tarter 2000) generated as beacons or for communication, the creation of a cosmic archaeological signature like a Dyson Sphere would not have required an active strategy on the part of the originating "civilization". An interesting distinction between systematic searches for objects like Dyson Spheres and SETI searches for radio and laser beacons is that for a Dyson Sphere search no presumption has to be made concerning the motivation of the originating civilization.

There could be several types of partial Dyson Spheres. A star could be surrounded by a uniform shroud that did not fully attenuate the stellar visible light. A second possibility is a star eclipsed by a circling ring that is totally opaque to visible light. It is difficult to distinguish this case from a pure Dyson Sphere. However the Dyson Sphere luminosity is reduced by, at minimum, two times the fractional coverage of the ring. The factor of two assumes that for small fractional coverage both the inside and outside of the ring radiate. In this article this possibility is subsumed in the pure Dyson Sphere signature. Note that a ring could be easier to construct than a shell but would also have less useful area and available energy. Dyson Sphere engineering is discussed in the appendix.

For a Dyson Sphere the stellar energy from the star would be reradiated at a lower temperature. If the visible light was totally absorbed by a thin "shell" a pure Dyson Sphere signature would be an infrared object with luminosity equivalent to the invisible star and a blackbody temperature corresponding to the radius of the spherical shell. For a sun-like star with the shell at the radius of the Earth the temperature would be approximately 300 ºK. Interestingly Dyson does not discuss the radiation distribution from his object. Dyson's approach is a more general perspective since among other complications the different parts of the reradiating swarm of objects in the shell could be at different distances from the star.

The apparently distinct signature of a Dyson Sphere would seem to make it an ideal candidate for cosmic archaeology. This article is devoted to a Dyson Sphere search using LRS, the Low Resolution Spectrometer on board IRAS, an infrared satellite that flew in 1983. IRAS identified 250,000 infrared point sources and scanned 96% of the sky. IRAS is described in more detail section 3. In many ways the IRAS database is ideal for a Dyson Sphere search because IRAS employed four filters centered at 12, 25, 60, and 100 μm and a spectrograph that together covered much of the radiation range emitted by a Dyson Sphere with a temperature between 100 and 600 °K.

Prior to the search reported here using the LRS spectrograph plus the IRAS filters a preliminary search was performed using only the filter information. For the preliminary search sources were retained only if the estimated temperatures lay between $150 \leq T_e \leq 500$ °K based on the 12, 25, and 60 μm filters. Sources were selected by several cuts including the temperature estimate and the requirement that FQUAL, the IRAS database flux quality factor, be greater than 1 for all of the 12, 25, and 60 filters. (In the database FQUAL(i) was set to 1 if there was only an upper limit in the ith filter because there was no "hours-confirming" observation, to 2 for moderate quality where at least one "hours-confirming observation was missing, or to 3 for high quality.) The preliminary search estimated 1 out of every 600 IRAS sources might pass a test making it a potential Dyson Sphere candidate. However the actual temperature difference distribution formed from the filter temperatures estimated from the ratios F[12]/F[25] and F[25]/F[[60] was flat so that a 3 σ peak in one bin might have required about 25 sources. This suggested that less than one in 10,000 of the IRAS sources could potentially be a Dyson Sphere. The filter technique suffered from a significant limitation because of the effects of zodiacal light and cirrus in the 60 and 100 μm bands. The presence of zodiacal or cirrus dust in these bands is a limitation on the use of all four filters. To overcome these problems the search reported here has used only the 12 and 25 μm filters coupled with the IRAS spectrograph.

Earlier searches for Dyson Spheres included a program by Jugaku & Nishimura (2003) to look for partial Dyson Spheres as well as several IRAS-based searches by Slysh (1985) and Timofeev et al. (2000) using all four IRAS filters.

The following sections discuss Dyson Sphere mimics, the IRAS infrared satellite as a Dyson Sphere search tool, the new search using the IRAS Low Resolution Spectrometer, a discussion of possible candidates, the relation to earlier Dyson Sphere and SETI searches, and concludes with remarks on the implications of the search for limits on Dyson Spheres and observations on the possibility of future investigations and searches. An appendix covers Dyson Sphere engineering.

## 2. DYSON SPHERE MIMICS

To identify a Dyson Sphere one has to rule out more conventional natural sources with similar signatures. On the other hand, it is conceivable that a real Dyson Sphere could have been classified as a more familiar object. A number of astronomical objects have infrared signatures somewhat like a Dyson Sphere. Indeed, the birth and death phases of many stars are associated with heavy dust clouds that can give rise to a black body signature.

Objects with resemblances to Dyson Spheres include stars with thick dust shells, regions of dust in a galaxy and very young stars that form in regions of dust, Mira variables, planetary nebulae, as well as other Asymptotic Giant Branch stars (AGB) and post AGB stars. The challenges of classification of such objects with mid-infrared signatures are laid out by Volk et al. (1992) in their discussion of a search for extreme carbon stars.

Typically at a late stage of a star's evolution the outer atmosphere can be blown away. Molecules such as SiO and hydroxyl ions (OH) form as the gas condenses and can give rise to natural OH masers. Often the dust cloud is not thick enough to hide the star. Many of these sources show silicate emission while others exhibit silicate absorption. They may also show pulsations. An OH/IR signal is evidence against a possible Dyson sphere.

Carbon stars are an important component of this class. Many but not all carbon stars have a prominent SiC emission line at about 11.3 μm. For example extreme C stars with thick envelopes may not have the feature (K. Volk et al. 1992). The line position, width, strength, temporal behavior, and perhaps asymmetry can give important information about the dust clouds surrounding the star (Thompson et al. 2006). For some cases the SiC line is non-existent or too small to be detected. Under those circumstances the carbon star might closely resemble a Dyson Sphere. Much of the appreciation of carbon stars as infrared objects arose after Dyson published the Dyson Sphere conjecture in 1960.

Typically galaxies detected by the IRAS satellite are relatively close and are not point sources. Even a nearby Dyson sphere should be a point source in an infrared telescope.

A Dyson Sphere candidate with a blackbody distribution can have several characteristics such as a blackbody temperature, the distance from our Sun, magnitude in the infrared, and variability. It may also have a stellar signature in the visible or near infrared. As seen in Volk, Kwok, and Woodsworth (1993) and Chen et al. (2001) some of the source types discussed above populate the same region of an infrared color-color plot as a Dyson sphere candidate would. Non-Dyson Sphere objects can be eliminated using discriminants like spectral lines in the infrared or radio regime, implausible blackbody temperatures, established classifications, and statistical departures from a blackbody distribution.

## 3. IRAS AS A DYSON SPHERE SEARCH TOOL

For a comprehensive Dyson Sphere search it is useful to survey a significant part of the sky as IRAS did. Some satellite-borne spectrographs like the IRS on the Spitzer have characteristically covered limited regions of the sky. An ideal Dyson Sphere search instrument would also be sensitive to temperatures ranging from roughly 100 to 600 ºK so it would need to cover wavelengths from 3 to 100 μm. This eliminates 2MASS and NICMOS on Hubble (Thompson et al. 1998) as principal search tools. Ground-based infrared telescopes have problems in the 10 to 100 μm regime because of high sky background. Finally good angular resolution would be helpful for a Dyson Sphere search to sort out associations with nearby stars.

The database from the IRAS satellite (Beichman 1987) is currently the best existing resource available to address these requirements, in part because IRAS was designed to study dust and dust is often a part of objects that resemble a Dyson Sphere. The IRAS primary mirror had a diameter of only 60 cm so that the diffraction limited circle that enclosed 80% of the energy had a radius of 25 arc seconds for the 25 μm filter. The IRAS supplement states that "…the in-scan pointing errors [for IRAS] were typically smaller than 5 arc seconds." Further "…the limiting flux density, away from confused regions of the sky, was about 0.5 Jy at 12, 25 and 60 μm and about 1.5 Jy at 100 μm for point sources…" The LRS spectrometer sensitivity extended below 4 Jy as measured in the 25 μm filter for the search reported here.

Of course data from other instruments can be used to rule out a Planck distribution hypothesis for an individual IRAS source. IRAS 16562-4256 is a case in point. The LRS distribution almost appears to have a 357 ºK Planck distribution although there is a hint of an 11.3 μm feature. Filter data from MSX, the Midcourse Space Experiment (Price et al. 2001), follows the LRS data but shows a rise in the 4 μm filters. The MIRS spectrographic data from the Japanese Infrared Telescope in Space (IRTS) mission in 1995 (Murakami 1996) is generally higher and shows strong features at 8 and 11 μm. Data from 2MASS (Skrutskie 2006) rises to meet the MIRS and MSX data in the 5 μm region. All of this additional information suggests a rapidly fluctuating source with lines that may come and go.

The 2MASS survey with nearly 500 M sources was more sensitive than IRAS and had much better angular resolution. However a blackbody source centered on the 12 μm IRAS filter must be at least 10 Jy to register in the 2MASS 2.17 μm filter, a factor of more than ten higher than the minimum IRAS sensitivity. The GLIMPSE Galactic Infrared Mid-Plane Survey (Benjamin et al. 2005) has used the Spitzer Infrared Array Camera (IRAC) to cover the galaxy from longitudes of ±65 deg and latitudes |b|<1 degrees (wider at the galactic center) using four filters at 3.6, 4.5, 5.8, 8.0 μm with a sensitivity down to the millijansky level. Use of GLIMPSE coupled with 2MASS for a Dyson Sphere survey is discussed later in the "future possibilities" section. If an IRAS source can be correlated with a 2MASS or Spitzer object, one can exploit their better angular resolution and pointing accuracy to establish correlations with other source information.

## 4. THE IRAS LOW RESOLUTION SPECTROMETER SEARCH

The Dyson Sphere search reported here has used the extended Calgary atlas for the IRAS Low Resolution Spectrometer (see below). The preliminary search noted earlier used the 12, 25, and 60 μm filters to do a blackbody fit. The requirement that FQUAL(60) > 1 was dropped for this new search to address the problem that higher temperature Dyson Sphere sources would give smaller values in the 60 μm filter. Using the LRS data also overcomes the impact of cirrus and zodiacal light on the 60 μm IRAS filter as well as the associated impossibility of fitting a Planck distribution with just two filter points and still obtaining a measure of the statistical significance.

*4.1 The Calgary LRS Sample*

The Calgary group has compiled a comprehensive atlas of 11224 sources containing a significant portion of the available useful IRAS Low Resolution Spectra (Kwok et al. 1997). From this sample 10982 sources satisfy the requirement FQUAL(12) and FQUAL(25) be greater than one.

The original IRAS LRS data set contains some problems that appear when the short ("blue") and the long ("red") wavelength spectrographic sets are compared (Cohen et al. 1992). The corrected values generated by the Calgary group were chosen rather than the original IRAS set. These earlier problems were due in part to the assumption "that Alpha Tau has a spectrum like a 10000 °K blackbody at these wavelengths which overlooks a SiO band in the spectrum near 8 microns."

The Calgary atlas assigns each LRS source to one of a number of categories on the basis of spectrographic features and the background continuum shape. Kwok et al. (1997) summarizes characteristics of the categories used for classification by Calgary including the ones that might contain Dyson Spheres. Parenthetically, the Calgary group has devised an extended classification system partly with the object of improving automatic classification and has considered broadening the data set further (Gupta et al. 2004). The search reported in the present publication uses the original Calgary set since it is readily available and has been used for a wide range of applications. The original IRAS LRS automated classification scheme for the atlas of the 5425 best LRS spectra is also feature-based. In practice it has not been used because it covers only half of the spectra in the Calgary set. Furthermore, the Calgary group has pointed out some problems with the original classification. An alternative classification system could be developed based on the type of source. This might be helpful but it is necessary to keep an open mind since one is looking for an unusual object that could be mimicked by more conventional objects and misclassified. Note that while the Calgary and original LRS catalogs are comprehensive, they are not necessarily the last word on IR spectra and identification since more data is now available from facilities such as the Spitzer Space Telescope and 2MASS.

The search discussed here used a sequential series of cuts to reduce the data sample by moving from a coarse to a finer grained net and ultimately selecting a set of objects that could be Dyson Sphere candidates. The cuts included temperature, Calgary class, IRAS object type (the IRAS variable IDTYPE-association with other catalogs, was limited to 0-no association, 2-stellar, 3-other catalogs, and 4-appearance in multiple catalogs), a qualitative scan of the spectral distribution to look for a Planck-like spectrum, lines, or large fluctuations, further cuts on established classifications, and ultimately limits on the least squares fluctuations. Factors giving rise to the fluctuations are discussed below in *4.5 Fitting to a Blackbody Distribution.* A source with another well-established classification was generally discarded unless there was some reason to doubt the classification. These cuts and scans are discussed in the following subsections and summarized in Table 1.

### *4.2 Temperature*

Dyson's original hypothesis envisioned a sphere constructed with a radius to the star corresponding to the temperature range where water is a liquid. A broader picture might include a temperature span that incorporates the range of temperatures over which some automata could operate. An initial temperature estimate for each IRAS source in the sample was made using the ratio of the 12 and 25 micron filters. The ratio of the two filters gives a putative Planck temperature based on the assumption that the spectral distribution of the source is a single black body convoluted with the filter responses. The points on the curve of temperature versus filter ratio have been fitted with an arbitrary fourth order polynomial to provide temperature estimators for the analysis. The initial temperature span for this search was set somewhat arbitrarily to be $100 \leq T1 \leq 600\ °K$ where T1 is the color temperature

$$T1 = c_0 + \sum_k c_k (f_{12} / f_{25})^k \tag{1}$$

based on the flux ratio $f_{12}/f_{25}$ where $f_{12}$ is the IRAS FLUX(12), $f_{25}$ is the FLUX(25), and $c_k$ are the fitting constants. The fitted constants for the curve were $c_0 = 98.0$, $c_1 = 433.1$, $c_2 = -386.0$, $c_3 = 240.2$, $c_4 = -47.0$. Applying these temperature cuts leaves a sample of 6521 sources.

### *4.3 Selection by Calgary Groups*

In an effort to limit the sample for individual source investigation the LRS spectra for an earlier sample of 387 sources selected using the 12, 25, and 60 μm filters were examined and compared to the Calgary IRAS classification. The spectra most similar to a Planck distribution fell in the C, F, H, P, and U groups. Carbon stars are typically classified as C. F objects typically have featureless spectra and are often oxygen or carbon-rich stars with small amounts of dust. H cases have a red continuum often with a 9.7 μm silicate or an 11.3 μm emission feature attributed to PAH (Polycyclic Aromatic Hydrocarbons). For the H cases the blackbody distribution usually has a relatively low temperature. Often H cases are planetary nebulae or are in $H_{II}$ regions. For the Calgary LRS IDTYPES = 0, 2 with temperatures below 200 °K the LRS spectrum rises smoothly with wavelength, that is the fluctuations are small. P cases can have a red continuum with a sharp rise at the blue end of the spectrum or either an 11.3 or 23 μm PAH emission feature. Initially the P group was not analyzed. Later the group was investigated further but eventually all of the P classification candidates were rejected. Sources in the U group have "unusual spectra showing a flat continuum" where the nature of the source is typically unknown.

The other categories were not investigated for various reasons. The A group sources had a 9.7 μm absorption feature. They are often oxygen-rich AGB stars surrounded by a thick cloud of dust. No A group sources qualified as interesting Planck spectra in the 387 source sample. E type stars are also oxygen-rich AGB stars but with a 9.7 μm emission feature. Again none qualified in the sample. Stars categorized as I had noisy or incomplete spectra. L type stars, a small group, had emission lines above a continuum. S types, another small group, had spectra that corresponded to a Rayleigh-Jeans optical tail.

Selecting on C, F, H, P, and U resulted in a sample of 2240 sources. The largest excluded group was E. It contained 3058 sources.

### 4.4 Selection by IRAS Object Type

The preliminary search focused on "pure" Dyson Spheres where there would be no visible source. Sample selection was done by taking objects with the IRAS variable IDTYPE equal to 0. Strictly IDTYPE = 0 corresponds to no catalog identification of the object. This requirement was later relaxed in two stages. In the first stage objects were included where IDTYPE = 2, an identified stellar object. After the IDTYPE = 0, 2 selection 1230 objects remained. Many sources in the 1230 sample were associated with near-by objects that are either visible or 2MASS stars. Once more it is emphasized that this approach allows the possibility of both pure and partial Dyson Spheres.

In a second stage sources with IDTYPE = 3, 4 were investigated. To facilitate processing only F and U sources were scanned. These categories contained essentially all of the interesting Dyson Sphere-like objects found in the IDTYPE = 1, 2 scan. This sample included 297 sources and resulted in an overall set with 1527 sources.

### 4.5 Fitting to a Blackbody Distribution

A large fraction of the 1527 source sample was fitted using a weighted and an unweighted least squares fit to a blackbody distribution. The weighted least squares form was

$$L(T, a_m) = \sqrt{\sum_i \left((P_{vi} - a_n \lambda_i C_i)^2 / P_{vi}\right) / 92} \qquad (2)$$

where $C_i$ are the so-called Calgary corrected spectral values corresponding to the raw LRS values in Watts/m$^2$/μm multiplied by the wavelength $\lambda_i$ and the Calgary λ-dependent correction, $a_n$ is a normalization factor introduced for graphical purposes, and $P_{vi}$ is proportional to the blackbody spectrum as a function of frequency (expressed in terms of $\lambda_i$, T) to fit the LRS spectral distribution

$$P_{vi}(a_m, T) = \left( \frac{2\pi hc a_m a_n}{\lambda_i^3 \left(e^{hc/\lambda_i kT} - 1\right)} \right). \qquad (3)$$

Here T is the blackbody temperature in Kelvin, $a_m$ is the fitting amplitude, h is Planck's constant, c is the speed of light, and k is Boltzmann's constant. The least squares form is essentially Pearson's $\chi^2$ statistic where the $P_{vi}$ weight the individual squares of the deviations. The least squares was minimized by adjusting T and $a_m$. (Strictly a value of 91 should have been used rather than 92 since there are two degrees of freedom for the fit.) An unweighted least squares was also calculated using

$$uL = \left( \sqrt{\sum_i (P_{vi} - a_n \lambda_i C_i)^2 / 92} \right) / N \qquad (4)$$

$$N = \sum P_{vi} / 92$$ .
(5)

In practice this was the more useful measure of significance. However, as noted, the weighted form was used for determining T and $a_m$. In the first portion of the cut sequence the least squares statistics were used to appraise but not reject sources.

Some consideration was given to the possibility of using the unweighted least squares to eliminate sources directly in the early part of the cut sequence. Deviations in the least squares arose from factors such as temporal variations in the source, mismatch between the "blue" and "red" elements of the LRS (actually two spectrometers with different resolutions), and rogue points that might have been generated in part by cosmic rays. Some of these features were apparent in the graphical display when comparing the red and blue parts of the LRS and the filter values. In short, the least squares was not just a statistical measure. On the other hand when the blackbody distribution least squares augmented with a Gaussian line was used to fit C stars there was no question that stars with distributions that contained lines 10 σ or more above a blackbody distribution had least squares below 0.1. Conversely, lines could still be seen down to 2-3 σ above a blackbody distribution. A second approach to using the least squares was to compare the blackbody fit (two parameters) to a second degree polynomial (three parameters). If they were approximately equal or the polynomial was a better fit then no clear case could be made for arguing the spectrum was a Planck distribution. This was often the case when the blackbody least squares was greater than 0.1. However for the initial scan and rescan sources were not eliminated from consideration based on the blackbody least squares.

*4.6 Direct Scanning*

Each source in the IDTYPE = 0, 2 sample was examined using the Strasbourg SIMBAD viewer to see if there was a nearby optical or 2MASS source. These cases were tagged but not eliminated. For the IDTYPE = 3, 4 set only the interesting cases were examined with the viewer. Typically, the spectrum for a source was reviewed by plotting the IRAS filter values (divided by the appropriate temperature adjusted Planck color correction from the IRAS Supplement) and the Calgary corrected spectroscopy data sets as well as the three filter values for the brightest/nearest 2MASS source. Direct scanning of the spectra for non-Planck shapes, obvious spectral lines, large data scatter, or other discriminants eliminated about 80% of the 1527 object sample. These sources were tagged as non-Dyson Sphere candidates. In a second pass the non-candidates with unweighted least squares values below 0.25 were rescanned again to confirm that sources had been eliminated in a consistent way.

Figure 1 (IRAS 17446-4048) illustrates a very common source type in the Calgary sample. These are sources with the 11.3 μm silicon carbide (SiC) emission feature typical of carbon stars. Often the background below the line is close to a blackbody spectrum. Both Calgary and SIMBAD classify 17446-4048 as a C star. The LRS spectral data are plotted in two sets, the short wavelengths (x) and the long wavelengths (dots). The filter values for F[12] and F[25] are shown as diamonds. The Planck distribution fit to the LRS including a Gaussian is plotted as a solid curve. The dotted curve is a fit to the corrected F[12], F[25], and F[60] filters. The 2MASS values are shown as triangles. Several filter points from the DIRBE experiment (Diffuse IR Background Explorer) on COBE are shown (open circles). These have large errors and an angular resolution of only 0.7

degrees but do suggest a blackbody distribution that extends below the LRS wavelengths. Notice that the 2MASS points are also consistent with the blackbody distribution. The spectrum here is fitted with a Gaussian centered at 11.3 μm with a width (σ) of 0.7 μm and an amplitude of 0.3 compared to the Planck distribution at 11.3 μm and a blackbody distribution with T = 594 ºK (with the line in the fit). The peak of the Gaussian line is 13 σ above the Planck distribution. In this search the possibility was initially held open that it might be argued that something like the 11.3 μm feature could have arisen out of the construction of a Dyson Sphere. However the strength of the line is consistent with a natural process. Parenthetically this bright infrared source is close to the direction of the galactic center and at a distance of 1.48 kpc, has an apparent bolometric magnitude of 5.72 (Groenewegen et al. 2002) and is likely quite variable (note that the filter points are substantially higher than the LRS). There is a faint visible star in the same direction.

Figure 2 for IRAS 19566+3423 shows a 10.2 μm fitted silicate absorption feature that sometimes occurs in OH/IR stars. SIMBAD classifies it as a star with an OH/IR envelope while the Calgary classification is U. There is no nearby visible star but there is a weak 2MASS source associated with it. The spectrum here is fitted with a Gaussian with an amplitude of 0.7 compared to the Planck distribution, a width (σ) of 0.9 μm, and a blackbody distribution with T = 246 ºK. The dip of the Gaussian line is also 13 σ below the Planck distribution. The source is near the galactic plane but away from the center. There is also an OH maser signal associated with the source.

Lines and absorption features like those in Figures 1 and 2 were identified based on peak amplitudes greater than 2 σ above background. This is a reasonable approach since a serious Dyson Sphere candidate would have to have a high statistical significance compared to any other possible explanation. Note that "line" here is used to specify a peak in the distribution that in some cases might be an amalgam of several lines.

After the direct scan examination 295 sources remained. About three quarters of these were eliminated because of a convincing identification of the source with a known object like a carbon star (14%, classif in Table 1), a nascent spectral line often coupled with a classification issue (56%), a fitted LRS temperature outside the original band (5%), or a non-Planck shape (0.5%, np in Table 1). The nine remaining H sources (3%) were also eliminated. An H source type typically has a low temperature (none of the nine remaining had temperatures above 180 ºK), is often in an $H_{II}$ region, and may have an associated OH maser. Of the five H cases with intriguing Planck spectra fits and relatively low unweighted least squares (IRAS 09032-3953, 13129-6211, 15103-5754, 17311-4924, 20028+3910) four had previously been identified as planetary nebulae or carried a GLMP planetary nebulae catalog identification (Garcia-Lario, Manchado, Pych, Pottasch 1997) associated with colors similar to planetary nebulae. Figure 3, IRAS 13129-6211, illustrates a case for the Calgary H classification. It may be near an $H_{II}$ region. The blackbody temperature is 132 ºK. In this case the source is near a visible star. Removal of all these sources left a sample of 65 possible candidates.

Larger least squares in the 65 source sample were loosely associated with high values in the F[100] filter (typically due to dust). More than 80% of the H (and P) sources were in bright sky regions with high F[100] values (F[100] > 100) while only 25% of the C and F sources had F[100] > 100. Several "bright sky" cuts were tried based on this but the cuts were either not effective or were unduly constricting on the search. This approach was even less effective for the F[60] filter. Similarly CIRR3, the "total surface brightness of the sky surrounding the source in a 1/2 degree beam at 100 microns," was not a useful discriminant. This was often high for sources close to the galactic plane so it excluded those sources rather than ones with poor fits.

The unweighted least squares values were used at this point to cut out some 43 sources because they had least squares values > 0.25. Typically for this situation the correlation coefficient for the Planck fit was below 0.6-0.7, it was difficult to unfold Gaussian lines, and a polynomial fit was often better than a Planck fit. Notice that these sources could still be Dyson Sphere candidates since they were ruled out because the available data from IRAS data was not strong enough statistically. Five more sources with unweighted least squares above 0.19 were eliminated because they had an association with a very bright 2MASS star (IRAS 06169-1235, 13553-5256, 16167-5118, 17096-3243, 21114+5013). A bright 2MASS source can give rise to a Rayleigh-Jeans tail that extends into the IRAS LRS region.

Finally one low least square case was eliminated because the source had previously been associated with a conjunction with an asteroid. This was found in the process of seeking an explanation for why no CO line appeared for this object when it was selected for observation in a group of putative carbon-rich AGB stars (Volk, K. et al., 1993). The spectral information for IRAS 19159+1556 is shown in Figure 4. There is a wide discrepancy between the spectrometer and the filter values and also MSX (open squares). The IRAS variability index is 99%. Further the F[100] filter is ten to twenty times higher than the other filters (but roughly consistent with a Planck fit to the LRS). The temperature is low, 208 ºK, suggesting that it perhaps should have been classified H rather than U. There is also no Spitzer information, perhaps because the source is too bright. (The source is not in the Spitzer GLIMPSE region.)

## 5. POSSIBLE CANDIDATES

This cut process left only sixteen candidates. Thirteen of these had least squares above 0.15. The three in the low least squares set are listed first in Table 2 ordered by the IRAS identification. The galactic coordinates are given after the IRAS identification. VAR, the IRAS percent likelihood of variability and IDTYPE are shown next. The presence of a visible star is indicated by one of three tags – bright, normal, or weak in the next column. This already subjective classification is handicapped for some crowded star fields. A quantitative estimate for any visible or infrared star associated with the object is given by the 2.2 μm $K_s$ filter value in magnitudes. An asterisk marks cases for $K_s$ shown in SIMBAD. The Calgary classification and fitted blackbody temperature follow. The fitted Planck distribution amplitude, $a_m$, and the unweighted least squares are shown next. After that there is a rating on the possible Dyson Sphere candidate (3 for nothing wrong, 2 for good but other classification, 1 for a problem, 0 for a significant problem). The Y axis in the spectral figures can be converted to Janskys by dividing by $F_{acc}$. The table also has notes on each source including the spectral classification where known. Interestingly 7 of the 16 sources were LRS spectra not included in the IRAS LRS Atlas set and two of these had low least squares.

Notice that only U and F sources remain. U sources had temperatures more along the line of the original Dyson suggestion for a temperature while the F source temperatures were higher. Most sources had a visible or 2MASS counterpart or both. In retrospect, focusing on just the F and U classifications would have resulted in a search sample one third smaller. What was gained by the search over other categories was more confidence that Dyson Sphere-like candidates were not being misclassified into these sets. This was particularly true for C cases which provided a yardstick on how small an amplitude a spectroscopic line could have had before it was lost in a blackbody distribution.

Figure 5 shows the spectrum of IRAS 20369+5131 in the low least squares set, the closest approximation to a Dyson Sphere signature in the sample. There is also confirming MSX infrared data shown as open squares for this source. For completeness several filter points from DIRBE are shown as open circles. At small wavelengths they lie well above the extrapolated blackbody distribution, MSX, and the closest 2MASS source. It may have been that the wide angular resolution of DIRBE overlapped a second source. There is no visible star and a null search (Deguchi et al. 2005) for the SiO maser J = 1 - 0 radio line. The blackbody temperature is 376 ºK. The Calgary classification is U. While this looks like a pure Dyson Sphere SIMBAD classifies it as a carbon star. Typically this classification would be based on other spectra in the millimeter regime. However no such information was found in a literature search. The source is slightly above the galactic plane and in a direction away from the galactic center. The height above the galactic plane may suggest it is relatively nearby.

Determining the distance of a featureless infrared source is difficult. It may be possible to estimate a kinematic distance for a partial Dyson Sphere or an associated source such as a binary companion. This could be estimated with a proper motion or a red shift (Jiang et al. 1996), perhaps from a maser signal. Two of the sixteen sources had measured proper motion error ellipses and ten of them had visible sources. No kinematic distance determination was made for IRAS 20369+5131. On the other hand Jiang et al. (1997) have estimated a distance of 4.63 kpc for another of the sixteen sources, IRAS 03078+6046. This source is discussed further below.

In an early IRAS-based Dyson Sphere search Slysh (1985) developed a bolometric formula for the peak flux density of a Dyson Sphere spectrum:

$$S_m(Jy) = \frac{35}{T}\frac{1}{D^2}\frac{L}{L_o}, \quad (6)$$

where D is the distance to the source in kpc and L and $L_o$ are the luminosities of the putative Dyson Sphere and the Sun. Notice that this is independent of the Dyson Sphere radius except for the fact that the temperature is a function of radius. This equation can be solved to get a bolometric distance for a Dyson Sphere:

$$D_{bol}(L) = \sqrt{\frac{35}{S_m T}\left(\frac{L}{L_o}\right)}. \quad (7)$$

For IRAS 20369+5131 $D_{bol}$ = 42 pc for a Dyson Sphere with the luminosity of the Sun.

For a partial Dyson "Sphere" due to a ring the bolometric distance is of the order of $D_r = D_{bol} \times \sqrt{2 \times f_c}$ where $f_c$ is the fractional areal coverage of the ring. For a ring that covers 0.5% of the sphere the effective bolometric distance will be one tenth of that for a full coverage sphere. Partial Dyson Spheres effectively need to be closer in order to be detected.

Another gauge of source distance is to assume that the source is at a scale height above or below the galactic plane and then use b, the galactic latitude, to calculate another distance, $D_{scl}$. Clearly this is not a measurement but a perspective on what the distance might be. Somewhat arbitrarily assuming a galactic scale height of 250 pc and a Sun height above the galactic plane of 34.5 pc gives $D_{scl}$ = 1970 pc for IRAS 20369+5131. The luminosity of a source at that distance would have to be L = 2160 $L_{sun}$. Slysh notes that this is typical of the luminosities of some late red giants. One is then left with two possibilities, a relatively nearby DS with a luminosity on the order of our sun or a distant and invisible red giant with no noticeable 11.3 μm emission peak (that is no peak above 2 σ).

IRAS 20331+4024 is a second source in the low least squares set. It is quite similar to IRAS 19159+1556, the source ruled out because of a conjunction with an asteroid. The F[60] and F[100] filters are very high and suggest an even lower temperature. The bolometric distance is 45 pc. IRAS 20035+3242 is a similar source with an even lower fitted temperature of 185 °K. The LRS distribution produces a ragged but plausible Planck fit and matches the information from MSX with a scale change. IRAS 11544-6408, classified as a post-AGB star by SIMBAD, is also similar but there are hints of a non-Planck distribution associated with a least squares > 0.19. A recently released Spitzer IRS measurement (Garcia-Lario campaign ID 1085) shows small but distinct features and a non-Planck distribution that follows the LRS spectrum. The fitted black body temperature is 211 °K.

Figure 6 (IRAS 16406-1406) is a third possible candidate in the low least squares set. The Calgary classification is F. The 2MASS points match up with the LRS Planck fit. In this case a polynomial fit is about as good as a blackbody distribution. There is nothing reported for MSX. Several filter points from DIRBE are shown as open circles. They suggest a blackbody distribution that extends below the LRS wavelengths. The fitted blackbody temperature is 538 °K. There is a faint hint of an emission artifact at 13.5 μm and a weak visible star. The source is in the direction of the galactic center but above the galactic plane. There is an indication of variability. With the available information it would be hard to make a case that this source is an interesting example of a possible Dyson Sphere candidate. The irregular distribution and possible line artifact further weaken this source as a candidate. The one solar bolometric distance is $D_{bol}$ = 55 pc. The galactic scale distance is 600 pc which would require a luminosity of 130 $L_{sun}$.

Figure 7 shows the spectrum for IRAS 03078+6046 in the high least squares set. This could be a Planck spectrum but the scatter in the points is high. This illustrates the problem of distinguishing a blackbody distribution when the deviations become larger. For example, a polynomial fit is slightly better than a blackbody distribution. The blackbody temperature is 381 °K. The IRAS variability index is 69% and the filter values are different than the LRS. Four points from MSX follow the distribution of the LRS but are 60% higher. Kerton & Brount (2003) have found no CO associations for this source. There is also no DIRBE data. The object is near the galactic plane but in a direction opposite to the galactic center. There is not enough information to make a strong case for the source as a possible Dyson Sphere candidate. The bolometric distance for a Dyson Sphere with the Sun's luminosity is $D_{bol}$ = 86 pc. The galactic scale distance is 4.8 kpc which would require a luminosity of 3000 $L_{sun}$. As noted earlier Jiang et al. (1997) give D = 4.63 kpc. In fact this source has been used to set $Z_{scl}$ and force rough agreement between the scale distance and the Jiang et al. determination. If the Jiang distance is used the large luminosity implies a late red giant rather than a Dyson Sphere (although there appears to be no visible star).

The data for IRAS 18094-1505 in the high least squares set is similar to IRAS 03078+6046 where the scatter in the LRS points is high. However the MSX data does follow the distribution of the LRS and the 2MASS points are roughly compatible. There is no DIRBE data. There is a visible star. The Planck temperature is 333 ºK. IRAS 18094–1505 is classified as a semi-regular pulsating variable star. The IRAS variability index is 8%. The object is near the galactic plane. Combining the LRS, 2MASS, and MSX data might make a statistically stronger case for the source as a possible Dyson Sphere candidate. The one solar bolometric distance for a Dyson Sphere is $D_{bol}$ = 84 pc. The galactic scale distance is 8 kpc which would require a luminosity of 9000 $L_{sun}$. IRAS 18209-2756 (291 ºK), IRAS 18287-1447 (386 ºK), and IRAS 18298-2026 (377 ºK) are somewhat similar to IRAS 18094-1505. All of them have larger fluctuations. In addition IRAS 18209-2756 has a very weak hint of a 12 µm emission line. IRAS 20212+4301 and IRAS 18013-2045 are also similar. Adding the MSX and 2MASS points suggest Planck temperatures greater than 600 ºK for these two. IRAS 18112-1353 resembles these but has also exhibited a large variation between the MSX, IRAS filters, and LRS data.

Finally, three of the sixteen sources, IRAS 00477-4900, 02566+2938, and 19405-7851 have prominent IR short wavelength features observed via 2MASS, MSX, and or DIRBE. Characteristically, the short wavelength behavior merges into a longer wave length tail that is only incidentally fitted when just the LRS data is used to match a Planck distribution. These would seem to be unlikely partial Dyson Sphere candidates since the dominant mechanism giving rise to the IR is producing the behavior in the 2MASS region.

The distribution of the sixteen sources is shown in Figure 8 on a galactic Aitoff plot. The three sources with smaller unweighted least squares values are plotted as squares, the others as circles. The 2240 source sample resulting from the selection on LRS cases, FQUAL(12) and FQUAL(25) > 1, temperature, Calgary ID types, and IDTYPE = 0, 2, 3, 4 is also plotted as dots. (For IDTYPE 3, 4 all Calgary classifications are shown, not just F and U). About three quarters of the 16 sources are near the galactic plane. Four others are away from the plane suggesting that they are relatively nearby. Note that the galactic plane is prominent in the 2240 source set. However there is not much sign of the galactic bulge. This suggests that the LRS might not reach out to the galactic center for this sample.

## 6. RELATION TO EARLIER DYSON SPHERE SEARCHES

A number of Dyson Sphere searches have been carried out over the years. Searches such as Jugaku et al. (2003) and unpublished searches conducted by Conroy, C. & Werthimer, D. and Globus, A. et al. (2003) were directed toward partial Dyson Sphere signatures. These groups looked for slight infrared excesses around visible stars. This approach is sound as a tool to look for cases where the fractional coverage of the Dyson Shroud is incomplete. Indeed, since pure Dyson Spheres may be hard to construct, partial Dyson Sphere searches could be the best attack in a search for examples of cosmic archaeology. This approach is not appropriate for a pure Dyson Sphere search because the host star would not be visible.

As noted earlier Slysh (1985) and TKP (Timofeev, Kardashev, & Promyslov 2000) carried out early searches for Dyson Spheres using IRAS. TKP used blackbody temperature fits to the four IRAS filters to select about 100 sources with temperatures in narrow intervals around 115 °K and 285 °K from the 3000 brightest IRAS sources. While the first phase of the program discussed here used something like this approach (three lowest filters rather than all four) the investigation reported in this article uses the LRS for the temperature fit. The TKP technique effectively excludes dusty regions where F[60] and F[100] are large. Of the 14 sources listed in the TKP tables and Figure 3 of their article, only six have LRS spectra and none of these were in the 1527 source sample investigated in this search. All six have Calgary classifications that would have excluded them from this search. For five of the six with LRS data the temperature fits from the LRS differ from the four filter fits by an average of 110 °K (the LRS temperatures are all higher). This illustrates that filter temperature fits can be misleading. Two of the six have FQ[12] values of 1 which would also have excluded them. Most are associated with definite stellar types.

Slysh reports on the early IRAS data release noting that there are many sources in the roughly 200,000 sample with 50 °K < T < 400 °K. He gives details on six sources found by employing four filter blackbody fits. Four of the sources had LRS information. It is not clear how large a fraction of the IRAS data set was searched to find these sources. The four filter temperatures reported in the paper range from 85 to 350 °K. One of the sources is in the 1527 source sample. That source, IRAS 04530+4427 (Slysh 0453+444), has a prominent 12 micron emission line and is classified as a carbon star (CGCS 6092). IRAS 05073+5248 (Slysh 0507+528) was excluded from the 1527 source set because of the E classification. It is associated with NV Aur, a Mira Cet that has an OH/IR maser associated with it. IRAS 05361+4644 (Slysh 0536+467) has an odd spectrum that includes a 12 micron peak. It was also classified as E by Calgary and therefore excluded. SIMBAD associates it with the carbon star BD+46 1033. IRAS 17411-3154 (Slysh G 357.3-1.3) has a strong absorption feature at 10 microns. It was classified A by Calgary and excluded. For the four with LRS data the temperature fits from the LRS differ from the four filter fits by an average of 44 °K (the LRS temperatures are all higher).

In summary none of the Slysh or TKP sources is included in the list of interesting sources reported here and only one is in the 1527 source sample. The examples from these papers demonstrate that the LRS is an effective tool in discriminating against filter spectra that only roughly follow a Planck distribution. The examples also suggest that relying on information from the 60 to 100 micron region may obscure an evaluation of a source in the Dyson Sphere regime covered by the F[12] and F[25] micron filters.

## 7. RELATION TO SETI SEARCHES

Over the last decades several wide-ranging searches have been made for radio ETI signals. In Project Phoenix (Tarter 2000) the SETI Institute and its collaborators have used the Arecibo 300 m diameter radio telescope as well as the 42 m dish at Green Bank and the 64 m Parkes telescope in Australia to look at on the order of 1000 individual targets within 250 light years principally selected for characteristics that might be hospitable for life. These included F, G (Sun-like), and K type stars. A region from -2º to +38º declination in celestial coordinates was searched with Arecibo with a sensitivity of O(1 Jy). (The band is shown in Figure 8.) Individual observations were typically for 140 - 280 s per channel. The system monitored the band from 1.2 to 3 GHz. This band includes the 21 cm hydrogen line "water hole." No robust ETI signals have been detected.

The sixteen sources listed in Table 2 were compared to the Phoenix target list. None of the sixteen is in the Phoenix list. This is not surprising since a pure Dyson Sphere is not like our Sun. Even the Dyson Sphere candidates here with an associated visible star have unusually high infrared luminosity and would not be confused with an ordinary sun.

The Berkeley SERENDIP search (D. Werthimer, private communication) that operates the SETI@home program has also used data from Arecibo. SERENDIP has run continuously for a commensal survey measurement with other Arecibo programs. Characteristically while one observer uses the Gregorian feed, SERENDIP collects data from a 1420 MHz feed located near the 400 MHz feed at the other end of the rotating platform suspended above the reflecting dish. This approach has resulted in a wider effective sky coverage than Phoenix at the expense of some loss of sensitivity compared to Phoenix. SERENDIP sampled over ~1 Hz windows in a 100 MHz frequency range centered around the 21 cm line. Two of the 16 sources discussed above were in the SERENDIP survey region (IRAS 02566+2938 and IRAS 20035+3242, both high least squares sources). No concrete SETI signatures were found in the course of SERENDIP so in that context these two IRAS sources are free of SETI-like signatures to the level of the search.

Somewhat similar programs have been conducted at Harvard. Likewise several optical searches have been carried out. None of these projects has detected a confirmed ETI signal. A SETI search of thirteen of the sixteen sources discussed in this article is planned for the Allen Telescope Array (ATA) in the near future. There is, of course, no particular reason to expect that a Dyson Sphere would be an ETI radio or optical emitter.

## 8. CONCLUDING REMARKS

This Dyson Sphere search has looked at a significant fraction of the IRAS LRS sources with temperatures under 600 ºK. Since IRAS covered 96% of the sky this is essentially a whole-sky search. Indeed this search may be one of the only SETI/cosmic archaeology whole-sky searches conducted so far. Unlike many radio and optical SETI searches this one does not require purposeful intent to communicate on the part of the originating source of the signature of intelligence.

The one sun bolometric distance, $D_{sol}$, depends on both T, the blackbody temperature, and $S_m$, the peak flux density. The largest $D_{sol}$ for the sixteen source sample was 118 pc for an $S_m$ of 8.6 Jy. Most of the 65 source sample had fluxes above 4 Jy which could give $D_{sol}$ values in the range 150 pc. The IRAS LRS reached down to filter fluxes of 1-2 Janskys in the F[12] and F[25] filters so that the IRAS LRS could in principle have found a Dyson Sphere with the luminosity of our sun out to 300 pc. This reach encompasses a region that contains on the order of a million solar type stars. Interestingly, if the effective galactic disk scale height is in the 100-200 pc range this view (that is, based on the one solar luminosity estimate) suggests that the sample of Dyson Spheres found with the IRAS LRS would be scattered more over the whole sky rather than lying near the galactic plane. As seen in Figure 8, perhaps one third of the potential candidate sources are scattered over the whole sky but the other two thirds are in the galactic plane. These sources are probably more luminous and at greater distances.

This search has shown that at best there are only a few quasi-plausible Dyson Sphere signatures out of the IRAS LRS sample in the $100 < T < 600$ °K temperature region. This limit includes both pure and partial Dyson Spheres. With several possible exceptions all the "good" sources identified in this search have some more conventional explanation other than as a Dyson Sphere candidate. In spite of the fact that there are many mimics such as stars in a late dusty phase of their evolution good Dyson Sphere candidates are quite rare!

*8.1 Possible Future Investigations and Searches*

More information on candidates in the sixteen source list could help to decisively rule them out or even strengthen a case for some of them as potential Dyson Spheres. Specifically observations of maser signals associated with a source should eliminate it as a candidate. Improved infrared observations could also eliminate sources.

The angular resolution for Spitzer is ten to twenty times better than IRAS while the sensitivity is three orders of magnitude greater. Could Spitzer/GLIMPSE data be used to substantially extend the search for Dyson Sphere candidates or to rule out candidates in the sixteen source list?

The GLIMPSE survey (Benjamin, et al. 2005) has used the Spitzer Infrared Array Camera (IRAC) to survey the galactic disk in a band of galactic longitude of +/-65 degrees and +/- 1 degrees in latitude. One third of the IRAS sources covered in this Dyson Sphere search lie in this region. In GLIMPSE this region contains 50 to 100 million sources, 90% of which are cool red giants. According to E. Churchwell (private communication) this is in line with an earlier model (Wainscoat, et al. 1992) of the infrared sky. The IRAS database contains roughly 80,000 sources in the GLIMPSE band so that one might aspire to collect a sample of potential candidates one thousand times larger, that is O(10,000) sources with some characteristics of a Dyson Sphere. From a different perspective since Spitzer is one thousand times more sensitive, the bolometric reach for a one sun equivalent would be past the center of the galaxy. Parenthetically, the Spitzer IRAC camera effectively saturates for the bright sources found by IRAS so that GLIMPSE is often silent on the sources surveyed here.

IRAC has filters centered at 3.63 µm, 4.53 µm, 5.78 µm, and 8.0 microns. A Planck distribution peak centered at 6 microns corresponds to a blackbody temperature of 850 °K. This is substantially higher than maximum temperature investigated in this Dyson Sphere search. Because of the filter range the IRAC filter set will see only the rising portion of the Planck distribution for a 300 °K Dyson Sphere. At 200 °K there might only be a signature in the 8 micron filter. The camera for the Japanese satellite Akari (formerly Astro-F) will provide all-sky, high spatial resolution information for 2–5, 5-12, and 10-25 µm filters in the future. A Planck distribution could be fitted to the filters, particularly if the 2MASS filters at 1.23, 1.66, and 2.16 µm were also used. For some sources additional spectrographic information from the Spitzer IRS spectrograph or

...

perhaps a ground-based instrument like Gemini could be used to define the Planck shape and rule out spectral features. The Spitzer GLIMPSE survey might be useful for sorting out the lower end of IRAS distributions, particularly to get information on the difficult 8 μm line region and also to give better information on source positions. Unfortunately only one of the sixteen candidate sources falls in the GLIMPSE region and there appears to be no information on it, possibly because the bright source has saturated the pixels in the region.

A second reason to be interested in the GLIMPSE study is that it could give a richer sample in the sky coverage region of the SERENDIP SETI survey. About twenty percent of the 2240 source sample was covered in the SERENDIP survey. Altogether 6% of the 2240 source sample reported in this paper is within the SERENDIP – GLIMPSE overlap region. Assuming the overall GLIMPSE survey could yield a thousand times more Dyson Sphere-like sources (~$10^4$ sources) then GLIMPSE might give about 70 with SERENDIP SETI nulls. This would codify a fraction for Dyson Sphere-like sources with no SETI signals at a certain level. It would not, however, advance the investigation of Dyson Sphere-like signatures.

ACKNOWLEDGEMENTS

The author would like to thank, J. Annis, C. Beichman, and K. Volk for guidance.

APPENDIX: DYSON SPHERE ENGINEERING

The premise of the search reported here was to look for Dyson Spheres based on the original idea put forward by Dyson including his addendum with additional information (Dyson 1960). This concept was a "loose collection or swarm of objects traveling on independent orbits…" Dyson's model leads to a sun where the optical radiation is effectively shielded by a shell-like shroud and the shell radiates somewhat like a blackbody with the radius of the shroud. As noted earlier, this is similar to a carbon star surrounded by a thick, low temperature dust cloud. The assumption of a fully shrouded body is restrictive and perhaps ingenuous. A partial Dyson Sphere would be a more practical object to build. The search described here initially started by ruling out infrared sources with associated visible stars. Later it turned out that allowing visible stars did not unduly complicate the search and those cases were included. In any case the original premise of a pure Dyson Sphere constitutes a rather clean signature. The spirit of the search was to use the signature without investigating the underlying hypothesis in detail. This too is ingenuous because it leads to a particular signature. It is worth reemphasizing that a number of other interesting searches have looked for partial Dyson Spheres. The following paragraphs discuss the scale of the construction of a Dyson Sphere as an engineering project and then go on to the costs to construct a "sphere", criticisms of the Dyson hypothesis, and the stability of the sphere.

The mass of a Jupiter-scale Dyson Sphere would be $2*10^{27}$ kg. Earth's biosphere is estimated to be in the $10^{15} - 10^{16}$ kg range. The largest ocean liner is $1.5*10^8$ kg while the immobile Great Wall of China is on the order of $10^{12}$ kg and took 2500 years to build. The International Space Station weighs $2*10^5$ kg so that it is a factor of ten billion trillion smaller than a Jupiter-class Dyson Sphere. In short, a Dyson Sphere is a large object! Dyson's perspective on this is to note that increases of scale such as this can be reached in 3000 to 4000 years based on a population growth of O(1%/year).

The energy cost of constructing a Dyson Sphere is high. In his paper Dyson notes that the energy to disassemble and rearrange a Jupiter size planet is equivalent to 800 years of total solar radiation, not the much smaller amount of solar radiation falling on the planet's surface. (The 800 year number is of the order of the gravitational self energy of Jupiter.) This poses a bootstrap problem in the building process. Initially the available power would be low but could rise as more surface area developed. In any case the construction time would probably be much longer than $10^3$ years. For scale, Earth-based projects such as cathedral or military fortification construction like the Great Wall of China have required times of $10^2$ to $10^3$ years. Mass loss time scales for circumstellar dust clouds can be in the $10^3$ to $10^5$ year range. Creation of these clouds requires a significant part of the stellar energy.

One of the criticisms of Dyson's proposal from the start has been based on the incorrect presumption that Dyson's object was a rigid, hollow sphere. Papagiannis (1985) has given a convincing calculation showing that a rigid spherical shell could not be built if only gravitational force is considered (that is, neglecting radiation pressure and centripetal effects). He then goes on to argue that a system of "space structures" would only obscure 1% of the light from the star. This is one of the justifications used for partial searches such as those of Jugaku. This argument would seem to ignore a shroud consisting principally of thin silicon chips, solar panels, or carbon nanotubes. Put differently, it is difficult to anticipate the size or the technology going in to the individual cells in a Dyson Shroud developed by an advanced intelligence.

Another criticism of the Dyson hypothesis says that a spherical swarm is not a stable system. Somewhat similar systems are studied in connection with the stability of elliptic and spiral galaxies as well as with planetary formation in the solar system. Presumably if instabilities develop in a rotating Dyson Sphere it could collapse to a disk under its own gravitational attraction. On the other hand the individual cells in the shell could be actively steered, perhaps by solar sails.

# FIGURE CAPTIONS & FIGURES

1. IRAS 17446-4048, a typical carbon star. The 11.3 μm SiC emission feature is characteristic of C stars. IRAS filter values - diamonds, DIRBE points – open circles, 2MASS – triangles, LRS (low) – x, LRS (high) – dot, fit to filter – dotted line, fit to LRS - solid. line (Scale: F(Jy) = y axis/0.0046)
2. IRAS 19566+3423 showing 10.2 μm silicate absorption feature occurring in OH/IR stars. The solid triangles are 2MASS points. (Same legend as Fig. 1, scale: F(Jy) = y axis/0.0103)
3. IRAS 13129-6211, a typical Calgary H type spectrum often associated with HII regions. (Same legend as Fig. 1, scale: F(Jy) = y axis/0.0051)
4. IRAS 19159+1556, a second interesting H-like source. Note wide deviation between the LRS and the IRAS and MSX filters. (Same legend as Fig. 1, scale: F(Jy) = y axis/0.0560)
5. IRAS 20369+5131, the closest approximation to a Dyson Sphere spectrum found in the search. MSX points – open squares. DIRBE points (open circles) do not agree with the 2MASS values and are possibly from another source. (Same legend as Fig. 1, scale: F(Jy) = y axis/0.0168)
6. IRAS 16406-1406, a third possible source but with a high deviation and a relatively high temperature. Note hint of an emission feature at 13.5 μm. (Same legend as Fig. 1, scale: F(Jy) = y axis/0.00425)
7. IRAS 03078+6046, a Planck-like spectrum with high deviation. It is hard to make a case for a source with fluctuations this high. (Same legend as Fig. 1, scale: F(Jy) = y axis/0.0541)
8. Galactic Aitoff plot for the sixteen sources. Three lowest least squares sources – squares, others - circles (one is shadowed). Sources in the 2240 source sample - dots. The SETI Arecibo region lies between the 38º and -2º bounds. The galactic plane is clear in the distribution but not the galactic bulge.

**TABLES**

Table 1
Source flow through cuts

| Selection | Sources |
|---|---|
| IRAS sample | 245,889 |
| Calgary LRS sample | 11224 |
| FQUAL(12), FQUAL(25)>1 | 10982 |
| $100 \leq$ Temperature $\leq 600$ ºK | 6521 |
| Selecting C[†], F, H*, P*, U | 2240 |
| IDTYPE = 0, 2, 3, 4 | 1527 |
| Possible sources by direct scan | 295 |
| -Line(166),classif(40),T(14),H(9),nonplnck(1) | 65 |
| Statistical uncertainty < 0.25 | 22 |
| Somewhat interesting sources | 16 |
| Most interesting but with questions | 3 |

* later eliminated, †mostly lines

<p style="text-align:center">Table 2<br/>Interesting sources</p>

| IRAS name | Galactic coor. longitude (mod 180) | Lat. | VAR | Visible star? | Ks (mag) | | Calg. class | T fit (K) | amin (LRS) | unLSQ | DS rating | Notes |
|---|---|---|---|---|---|---|---|---|---|---|---|---|
| 16406-1406 | 4.10 | 20.20 | 74 | | 8.7 | * | F | 538 | 2.353 | 0.096 | 3 | Hint of 13.5 μm |
| 20331+4024 | 79.83 | 0.10 | 33 | | | | U | 177 | 294 | 0.066 | 3 | 8 μm?, H?, MSX, fil ≠ LRS |
| 20369+5131 | 89.09 | 6.29 | 43 | | 7.9 | | U | 376 | 16.42 | 0.072 | 2 | C* but no reference? |
| 00477-4900 | -56.44 | -68.40 | 4 | bright | 1.862 | * | F | 550 | 1.64 | 0.159 | 1 | Prominent 2MASS, DIRBE |
| 02566+2938 | 153.54 | -25.31 | 2 | normal | 2.07 | * | F | 429 | 3.26 | 0.188 | 2 | M7, brt 2MASS |
| 03078+6046 | 139.17 | 2.59 | 69 | | 11.96 | | F | 381 | 3.742 | 0.163 | 3 | MSX > LRS |
| 11544-6408 | -62.91 | -2.17 | 21 | | | | U | 211 | 39.1 | 0.191 | 1 | post AGB, PN?,8 μm? |
| 18013-2045 | 9.21 | 0.47 | 93 | normal | 5.18 | * | U | 330 | 7.66 | 0.208 | 0 | 2MASS means higher T? |
| 18094-1505 | 15.08 | 1.56 | 8 | normal | 6.85 | * | F | 333 | 6.84 | 0.208 | 3 | 2MASS means higher T? |
| 18112-1353 | 16.34 | 1.77 | 99 | weak | 6.81 | | F | 387 | 3.48 | 0.218 | 3 | MSX > LRS |
| 18209-2756 | 5.03 | -6.88 | 99 | | 7.94 | | U | 291 | 5.91 | 0.246 | 3 | Weak hint of 12 μm |
| 18287-1447 | 17.56 | -2.39 | 20 | normal | 9.42 | | F | 386 | 5.65 | 0.195 | 3 | MSX < LRS |
| 18298-2026 | 12.66 | -5.25 | 35 | | 11.02 | * | F | 377 | 3.602 | 0.249 | 3 | |
| 19405-7851 | -44.51 | -29.33 | 4 | bright | 1.96 | * | F | 518 | 2.17 | 0.154 | 2 | Prominent 2MASS, DIRBE |
| 20035+3242 | 70.13 | 0.57 | 11 | weak | 11.92 | | U | 185 | 46.03 | 0.176 | 2 | Peculiar star |
| 20212+4301 | 80.64 | 3.42 | 7 | normal | 5.76 | | F | 412 | 3.06 | 0.246 | 1 | MSX, 2MASS: T > 600 °K? |

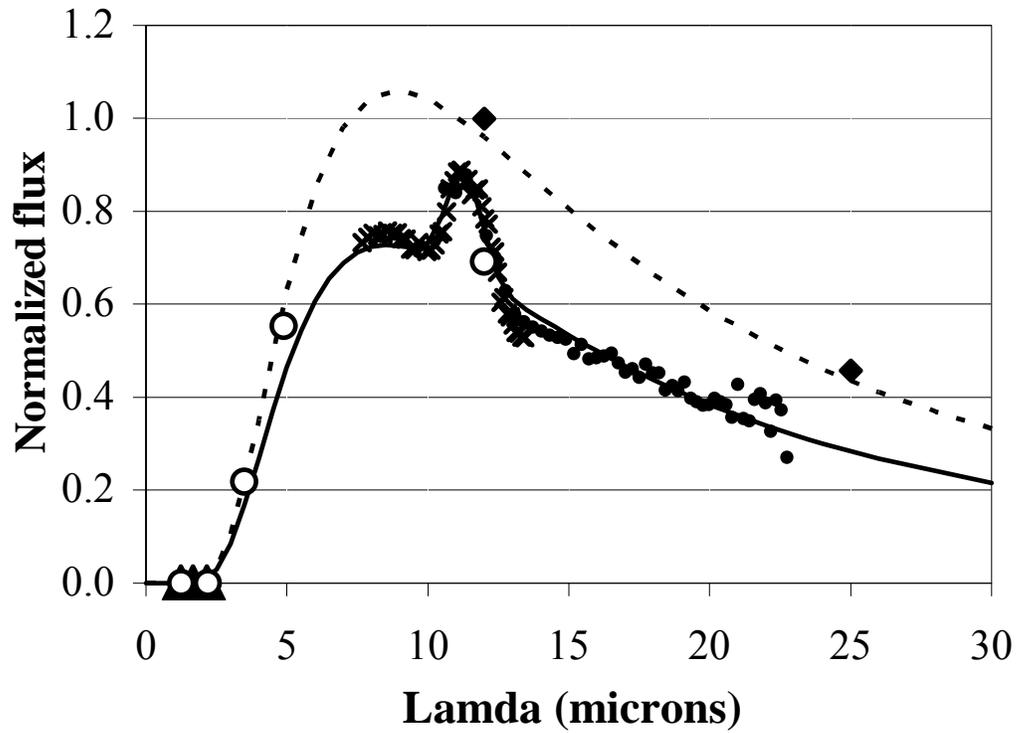

**Fig. 1: IRAS 17446-4048**

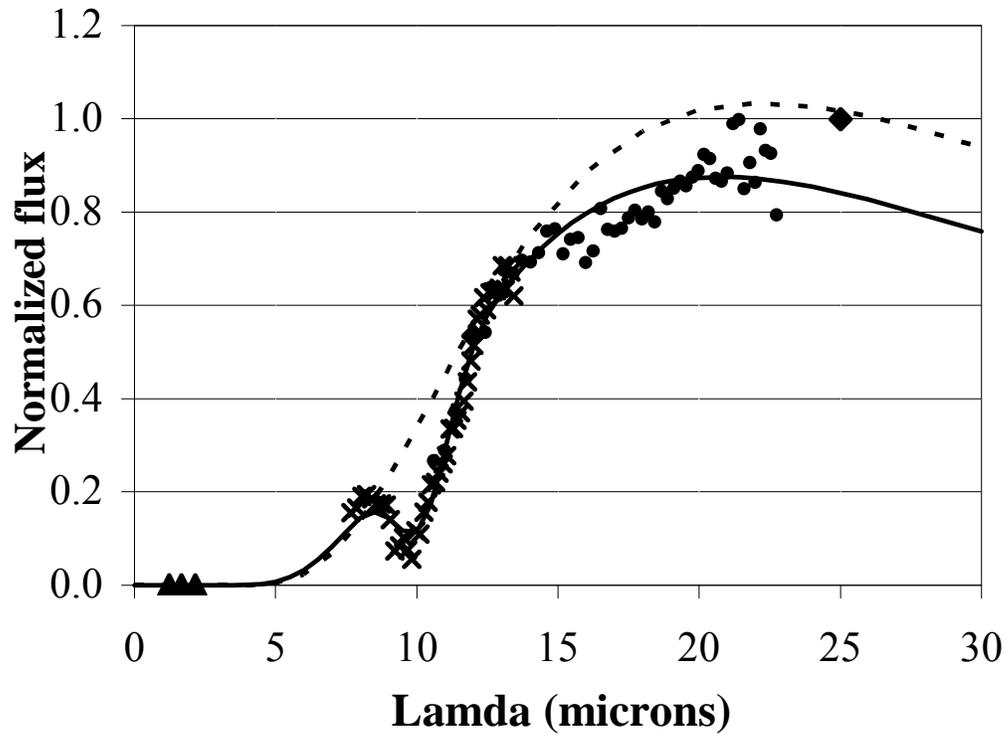

**Fig. 2: IRAS 19566+3423**

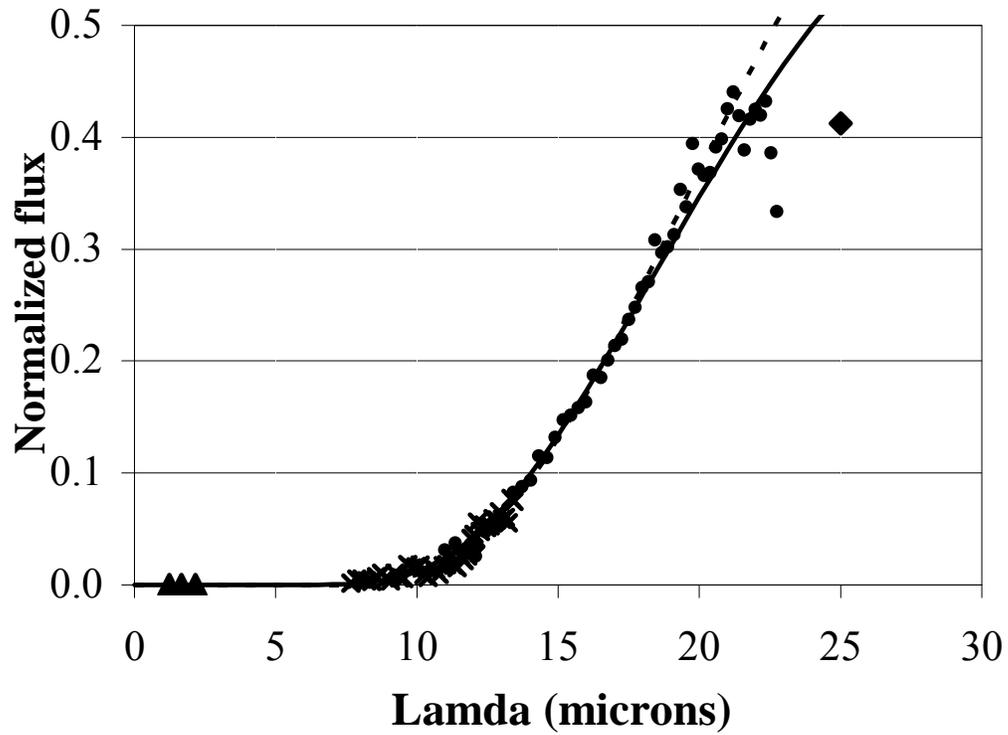

**Fig. 3: IRAS 13129-6211**

Figure 4

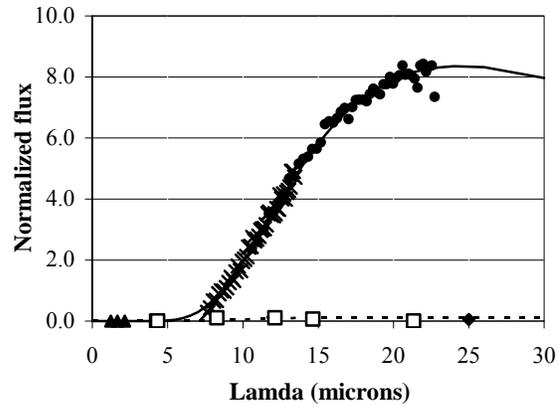

Figure 5

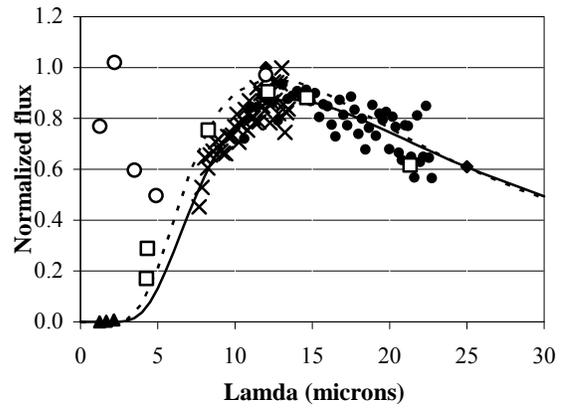

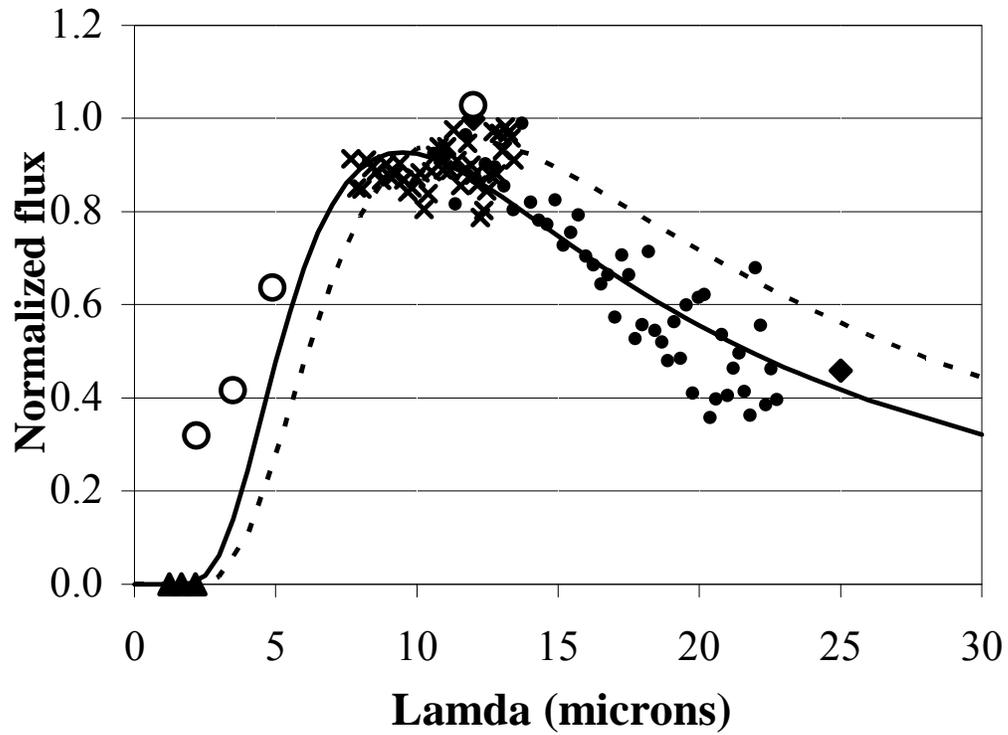

**Fig. 6: IRAS 16406-1406**

**Fig. 7: IRAS 03078+6046**

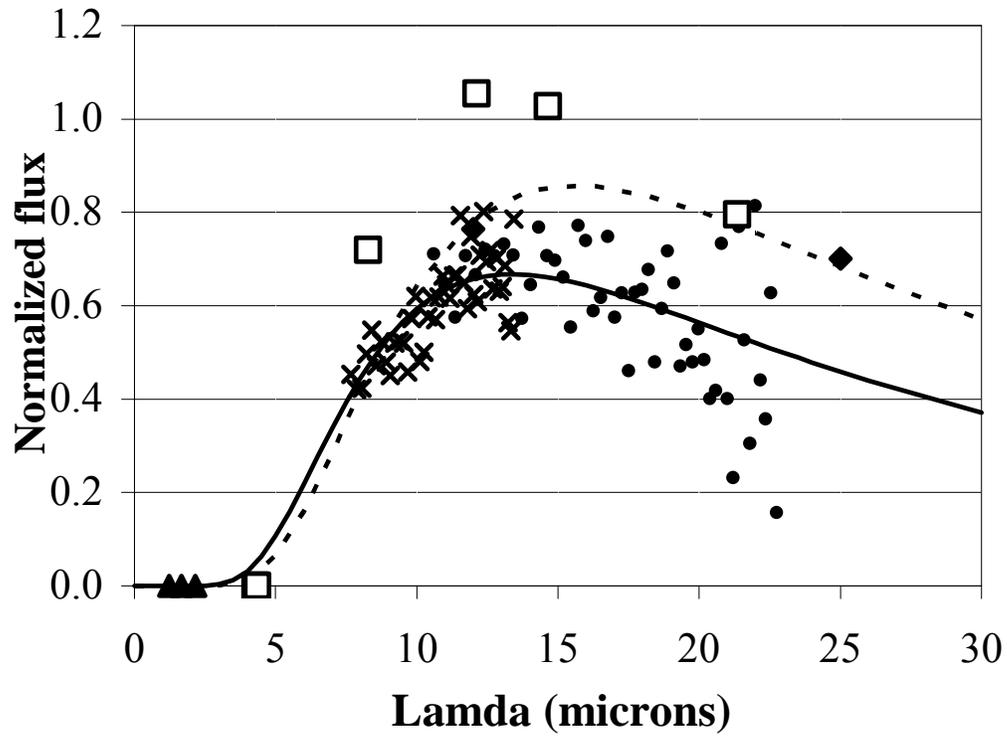

Figure 8

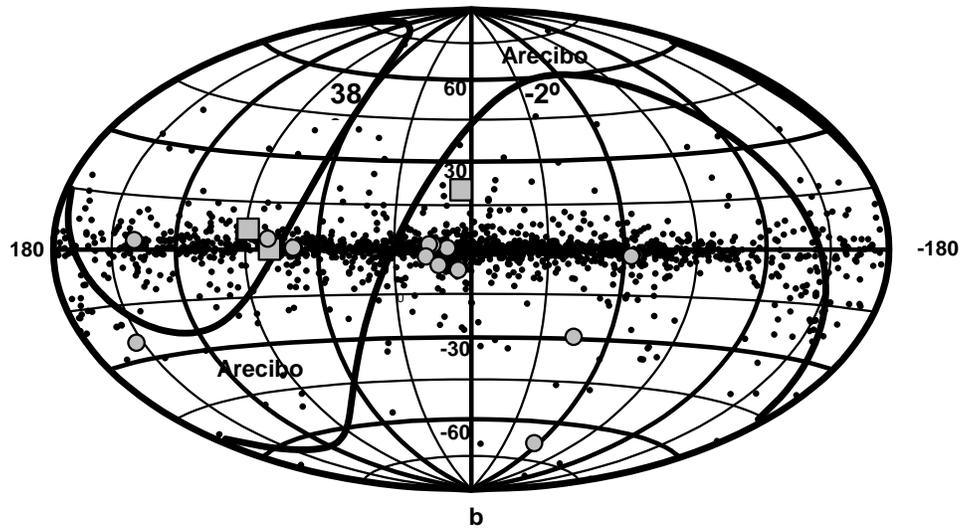